\documentclass{svjour3}                     % onecolumn (standard format)
\smartqed  % flush right qed marks, e.g. at end of proof
\usepackage{graphicx}
\journalname{Few-Body Systems (FB20)}
\begin{document}

\title{
%Insert your title here
Final state strong interaction constraints on weak \mbox{\boldmath $D^0 \to K^0_S \pi^+ \pi^-$} decay amplitudes
%Template for Proceedings of the Few-Body 20 Conference 
\thanks{Presented at the 20th International IUPAP Conference on Few-Body Problems in Physics, 20 - 25 August, 2012, Fukuoka, Japan}
}
%\subtitle{
%Do you have a subtitle?\\ If so, write it here
%Subtitle
%}


\author{B. Loiseau         \and
        J.-P. Dedonder \and A. Furman \and R. Kami\'nski \and L. Le\'sniak%etc.
}


\institute{B. Loiseau,   J.-P. Dedonder 
\at
              Laboratoire de Physique Nucl\'eaire et de Hautes \'Energies, Groupe Th\'eorie, 
Universit\'e Pierre et Marie Curie et Universit\'e Paris-Diderot, IN2P3 et CNRS, 4 place Jussieu, 75252 Paris, France \\
              \email{loiseau@lpnhe.in2p3.fr}           %  \\
        \and
         A. Furman    \at 
ul. Bronowicka 85/26, 30-091 Krak\'ow, Poland
        \and
         R. Kami\'nski, L. Le\'sniak   \at 
Division of Theoretical Physics, The Henryk Niewodnicza\'nski Institute of Nuclear Physics,
                  Polish Academy of Sciences, 31-342 Krak\'ow, Poland
}

\date{Received: date / Accepted: date}

\maketitle

\begin{abstract}
Weak decay tree and annihilation - t-channel $W$- exchange amplitudes  for the $D^0 \to K^0_S \pi^+ \pi^-$ process are calculated using quasi two-body QCD factorization approach and unitarity constraints.
 Final state strong $K \pi$ and $\pi \pi$ interactions in $S$, $P$ and $D$ waves are described through corresponding form factors including many resonances.
Preliminary results compare well with the effective mass distributions of the Belle and BABAR Collaboration analyses.

%Please provide an abstract of 100 to 150 words. The abstract should not contain any undefined abbreviations or unspecified references.
%Insert your abstract here. Include keywords, PACS and mathematical
%subject classification numbers as needed.



\end{abstract}

\section{Introduction}
\label{intro}

Why should one study the weak decays $D^0 \to K^0_S \pi^+ \pi^-$?
First, %this is a  self-conjugate reaction as $K^0_S=(K^0+\bar K^0)/\sqrt{2}$ 
%and 
the recent measurements of the $D^0$-$\bar D^0$ mixing parameters for this self-conjugate reaction by the BABAR~\cite{SanchezPRL105} and Belle~\cite{ZhangPRL99} Collaborations could show the presence of new physics contribution beyond the standard model.
Second, the Cabibbo-Kobayashi-Maskawa, CKM, angle $\gamma$ can be evaluated from the analyses of the  $B^\pm \to D^0 K^\pm, D^0 \to K^0_S \pi^+ \pi^-$ decays. 
Third, one can learn about the final state meson-meson interactions,  the  meson resonances decaying into different meson-meson pairs and their interferences in the Dalitz plot.
One can also perform a partial wave analysis of decay amplitudes.
In addition, constraints from quasi two-body QCD factorization~\cite{BenekeNPB675}, QCDF, approach will allow to test theoretical models of form factors entering in the decay amplitudes.
\paragraph{Quasi two-body factorization}
Following a program devoted to the understanding of the rare three-body $B$ decays (see e.g. Ref.~\cite{DedonderAPPB42}) the presently available  $D^0 \to K^0_S \pi^+ \pi^-$ data are analyzed in the framework of QCDF.
Neglecting the small $CP$ violation in $K^0$ decays, it will be assumed that $\vert K^0_S \rangle =( \vert K^0\rangle+ \vert \bar K^0\rangle)/\sqrt{2}$.
The three-meson final states $\bar K^0 \pi^+ \pi^-$ are approximated (quasi two-body approximation) as being formed by a meson-meson state,   $[\bar K^0 \pi^+]_{S, P, D}$ or $[\bar K^0 \pi^-]_{S, P, D}$ or $[\pi^+{\pi^-}]_{S, P, D}$ in a $S, P$ or $D$ wave created by a $q \bar q$ pair and  the remaining meson, $\pi^-$, $\pi^+$ or $\bar K^0$, respectively.
Amplitudes are derived from the weak effective Hamiltonian, $H_{eff}$ which is
a superposition of left-handed quark current-current operators.
For instance one has the operator $O_1=j_1 \otimes j_2$ where $j_1 =\bar s_\alpha\gamma^\nu(1-\gamma^5)c_\alpha\equiv (\bar s c)_{V-A}$ and
$j_2= \bar u_\beta\gamma_\nu(1-\gamma_5)d_\beta\equiv (\bar u d)_{V-A}$, $\alpha$ and $ \beta$ being color indices.
Applying QCDF to the part of the amplitude proportional to the product of the CKM quark mixing matrix elements $V^*_{cs}V_{ud}\equiv \Lambda_1$ one has,

\begin{eqnarray}
\label{AmpliL1}
    &&\left \langle \bar K^0 \ \pi^- \pi^+ \vert \ H_{eff}\vert D^0 \right \rangle  \simeq \frac{G_F}{\sqrt{2} } \ \Lambda_1 \Big \{
     a_1\left \langle \pi^+\vert j_2\vert 0 \right \rangle  \left \langle  [\bar K^0\ \pi^- ]_{S,P,D}\vert j_1 \vert D^0 \right  \rangle \nonumber \\
&&+\  a_2 \left \langle \bar K^0 \vert  j'_2 \vert 0 \right \rangle \left \langle  [\pi^+ \pi^-]_{S,P,D} \vert j'_1\vert D^0 \right \rangle 
 +\ a_2  \left \langle  0 \vert j'_1\vert D^0 \right \rangle \left [ \left \langle  [\bar K^0\ \pi^- ]_{S,P,D} \pi^+ \vert j'_2 \vert 0 \right  \rangle \right. \nonumber \\
&&+ \left. \left \langle  \bar K^0\ [\pi^+\pi^- ]_{S,P,D}\  \vert j'_2 \vert 0 \right  \rangle \right] \Big \}
\end{eqnarray} 
 where $G_F=1.166\times 10^{-5}$ GeV$^{-2}$ is the Fermi coupling constant, $a_{1,2}$ are effective QCD Wilson coefficients and $j'_1=(\bar u c)_{V-A}$,  $j'_2=(\bar s d)_{V-A}$ derive from $j_{1,2}$ via a Fierz transformation.
In Eq.~(\ref{AmpliL1}) it appears that the cross product of bilinear quark currents in $O_{1}$ factorizes, after the introduction of the vacuum state, into the product of two matrix elements.
The first ones are proportional to the $\pi^+, \bar K^0, D^0$ 
decay constants $f_\pi, f_K, f_{D^0}$ since $\left \langle \pi^+ \vert j_2 \vert 0\right  \rangle =i{f_\pi} p_{\pi^+},\left  \langle \bar K^0 \vert j'_2 \vert 0\right  \rangle =i{f_K} p_{K^0}, \left \langle 0 \vert j'_1 \vert D^0\right  \rangle =-i{f_{D^0}} p_{D^0}$, $p_{\pi^+}, p_{K^0}$ and $p_{D^0}$ being the $\pi^+, \bar K^0$ and $D^0$ four momenta, respectively.
The second ones are transition matrix elements or form factors as 
with $M_i= \bar K^0, \pi^\pm, \ j=j_{1,2}\ {\rm or\ } j'_{1,2}$ one has,
$\left \langle  [M_1M_2 ]_{S,P,D}\ \vert j \vert D^0 \right  \rangle$
$=\left \langle  \bar D^0 [M_1M_2 ]_{S,P,D}\ \vert j\vert 0 \right  \rangle$ and $\left \langle  [M_1M_2 ]_{S,P,D} \ M_3\ \vert j \vert 0 \right  \rangle$
$=\left \langle  [M_1M_2 ]_{S,P,D}\ \vert j \vert \bar M_3 \right  \rangle$.
It can be shown from field theory and using dispersion relations~\cite{Barton65} that these form factors can be calculated exactly if one knows the $D^0$-$[M_1M_2 ]_{S,P,D}$ or  $M_3$-$[M_1M_2 ]_{S,P,D}$ strong interactions at all energies. 

\section{Decay amplitudes}
\label{amplitudes}
\paragraph{Different type of amplitudes} Amplitudes with $c \to su\bar d$ transition  are proportional to $V^*_{cs}V_{ud}$ where $V_{cs} \approx V_{ud} \approx \cos{\theta_C} \approx 0.975$, $\theta_C$ being the Cabibbo angle.
There are~7  such allowed tree amplitudes: 3 for the $\pi^+ [\bar K^0 \pi^-]_{S,P,D}$, 3 for the $\bar K^0[\pi^+ \pi^-]_{S,P,D}$ and 1 for the $\bar K^0 \omega(\omega\to [\pi^+ \pi^-]_P$ by $G$-parity violation) final states.
Amplitudes with $c \to du\bar s$ transition, proportional to $\sin^2{\theta_C}\approx (0.225)^2$, are doubly Cabibbo suppressed. There are 6 of them as the $W$ meson cannot couple to the $[K^0 \pi^+]_{D}$ final state.
There are also 7 allowed tree annihilation or $t$-channel $W$-exchange amplitudes corresponding to the $c \bar u$ annihilation into $s \bar d$ and 7 doubly Cabibbo suppressed annihilation amplitudes from the $c \bar u$ annihilation into $d \bar s$.
Altogether, the quasi two-body QCDF approach leads to 27 non-zero - 13 tree and 14 annihilation - amplitudes, 

\paragraph{Transition matrix elements} Several meson resonances can decay into the  two meson final states in $S ,P$ or $D$  wave.
For the kaon-pion subsystems, the scalar resonances,
 $K_0^*(800)^\pm$ or $\kappa^\pm$ and  $K_0^*(1430)^\pm$ decay into $[\bar K^0 \pi^\pm]_S$, the vector resonances, $K^*(892)^\pm$,  $K_1(1410)^\pm$, $K^*(1680)^\pm$ into $[\bar K^0 \pi^\pm]_P$ and the tensor resonances $K_2^*(1430)^\pm$ into $[\bar K^0 \pi^\pm]_D$.
For the pion-pion subsystems, the scalar $f_0(600)$ or $\sigma$, $f_0(980)$, $f_0(1400)$ decay into $[\pi^+ \pi^-]_S$, the vector, $\rho(770)^0$, $\omega(782)$, $\rho(1450)^0$, $\rho(1700)^0$ into
 $[\pi^+ \pi^-]_P$ and the tensor $f_2(1270)$ into $[\pi^+ \pi^-]_D$.
This leads to a rich interference pattern in the Dalitz plot.
These resonances are used to write the three-meson transition to the vacuum as,  

 \begin{equation}
\label{transh1h2h30}
\left \langle M_1(p_1)[M_2(p_2) M_3(p_3)]_{S,P,D}\vert \ j'\vert 0\right \rangle \simeq G_{R_{S,P,D}^{M_2M_3}}(s_{23}) \left \langle M_1(p_1)R_{S,P,D}^{M_2M_3}\vert \ j'\vert 0 \right \rangle,
\end{equation}
where $s_{23}=(p_2+p_3)^2$ .  The vertex functions $G_{R_{S,P,D}^{M_2M_3}}(s_{23})$ describe the $R_{S,P,D}^{M_2 M_3}$ resonance decays into the states $[M_2 M_3]_{S,P,D}$.
A similar equation holds~\cite{LesniakMeson2012}  for the $D^0$ transitions [see Eq.~(\ref{AmpliL1})] to two-meson states.
As an example of application of Eq.~(\ref{transh1h2h30}) one can choose $M_1 \equiv \bar K^0$, $[M_2 M_3]_S\equiv [\pi^+\pi^-]_S$ and ${R_S^{\pi^+\pi^-}\equiv f_0(980)}$, so that,

\begin{equation}
\label{K0f0}
  \left \langle \bar K^0(p_1) f_0(p_2+p_3)\vert (\bar s d)_{V-A}\vert 0\right \rangle=-i\frac{m_{K^0}^2-s_{23}}{p_{D^0}^2} \ p_{D^0} {F_0^{\bar K^0 f_0(980)}(m_{D^0}^2)}+  1\ {\rm term},
\end{equation}
 where $p_{D^0}=p_1+p_2+p_3$. 
 The $\bar K^0$ to $f_0$ scalar transition form factor, $F_0^{\bar K^0f_0}(m_{D^0}^2)$, related to the $\bar K^0 f_0$ interaction at $(p_K+p_{f_0})^2=m_{D^0}^2$, is a complex number to be fitted.
The extra ``$1\ {\rm term}$" gives a null contribution when multiplied, in Eq.~\ref{K0f0}, by  $\left \langle 0 \vert \ (\bar u c)_{V-A}\vert D^0 \right \rangle$  [see Eq.~(\ref{AmpliL1})].
The vertex function $G_{f_0(980)}(s_{23}) \simeq \chi_2 \ { F_0^{\pi^+\pi^-}(s_{23})}$ where
${\chi_2}$ is a constant fitted to data and $F_0^{\pi^+ \pi^-}(s_{23})$ is the complex pion scalar form factor~\cite{DedonderAPPB42} related to $\left \langle [\pi^+ \pi^-]_S\vert \ (\bar u u)_{V-A}\vert 0 \right \rangle$.
The other vertex functions for the kaon-pion~\cite{El-BennichPRD79} or pion-pion~\cite{DedonderAPPB42} systems in Eq.~(\ref{transh1h2h30}) are described in terms of the corresponding form factors.
The $D$-wave meson-meson form factors are represented by relativistic Breit-Wigner formulae.
In the transition form factors for $R_{S,P,D}^{M_2 M_3}$  we choose $R_S^{K^0\pi} \equiv K^*_0(1430)$, $R_P^{K^0\pi} \equiv K^*(892)$, $R_S^{\pi\pi} \equiv f_0(980)$ and $R_P^{\pi\pi} \equiv \rho(770)^0$.

\paragraph{A selected amplitude}
From Eqs.~(\ref{AmpliL1}) and (\ref{K0f0}) one obtains for the $D^0 \to K^0_S \pi^+ \pi^-$  $S$-wave annihilation amplitude with $[\pi^+\pi^-]_{S}$ subsystem in the final state,

\begin{equation}
\label{AnS}
A_{n\ 2S} =-\frac{G_F}{2} \ \Lambda_1 \ \chi_2\ a_2  \ f_{D^0}  (m_{K}^2-s_0) \   F_0^{K^0 f_0}(m_{D^0}^2)\ F_0^{\pi^+ \pi^-}(s_0),
\end{equation}
where $s_\pm=(p_{\pi^\pm}+p_{K^0})^2, s_0=(p_{\pi+}+p_{\pi^-})^2$.
Other amplitudes can be derived similarly, their expression will be given in a forthcoming paper, however some explicit formulae can be found in Ref.~\cite{LesniakMeson2012}.

\section{Experimental $D^0 \to K^0_S \pi^+ \pi^-$ data}
\label{data}
\paragraph{Isobar model and unitariy} 
The experimental Dalitz plot analyses~\cite{SanchezPRL105,ZhangPRL99} are performed within the isobar model to describe the final state meson-meson interactions and
many free parameters are used (of the order of  2 per amplitude): the BABAR Collaboration model relies on 43 parameters and that of Belle on 40.
Amplitudes are not unitary neither in the 3-body channels nor in the 2-body sub-channels.
Two-body unitarity is incorporated in the present model: unitary form factors are used in the $K \pi$  $S$- and  $P$-wave~\cite{El-BennichPRD79} 
and in the $\pi \pi$  $S$-wave~\cite{DedonderAPPB42} amplitudes.
The branching fraction of the sum of these amplitudes is larger than 80\% of the total  $D^0 \to K^0_S \pi^+ \pi^-$ branching fraction.

\section{Preliminary results and concluding remarks}
\label{conclusion}
\begin{figure}
%  \includegraphics{example.eps}
 %\includegraphics{fb20_logo.eps}
\begin{center}
\includegraphics[scale=0.40]{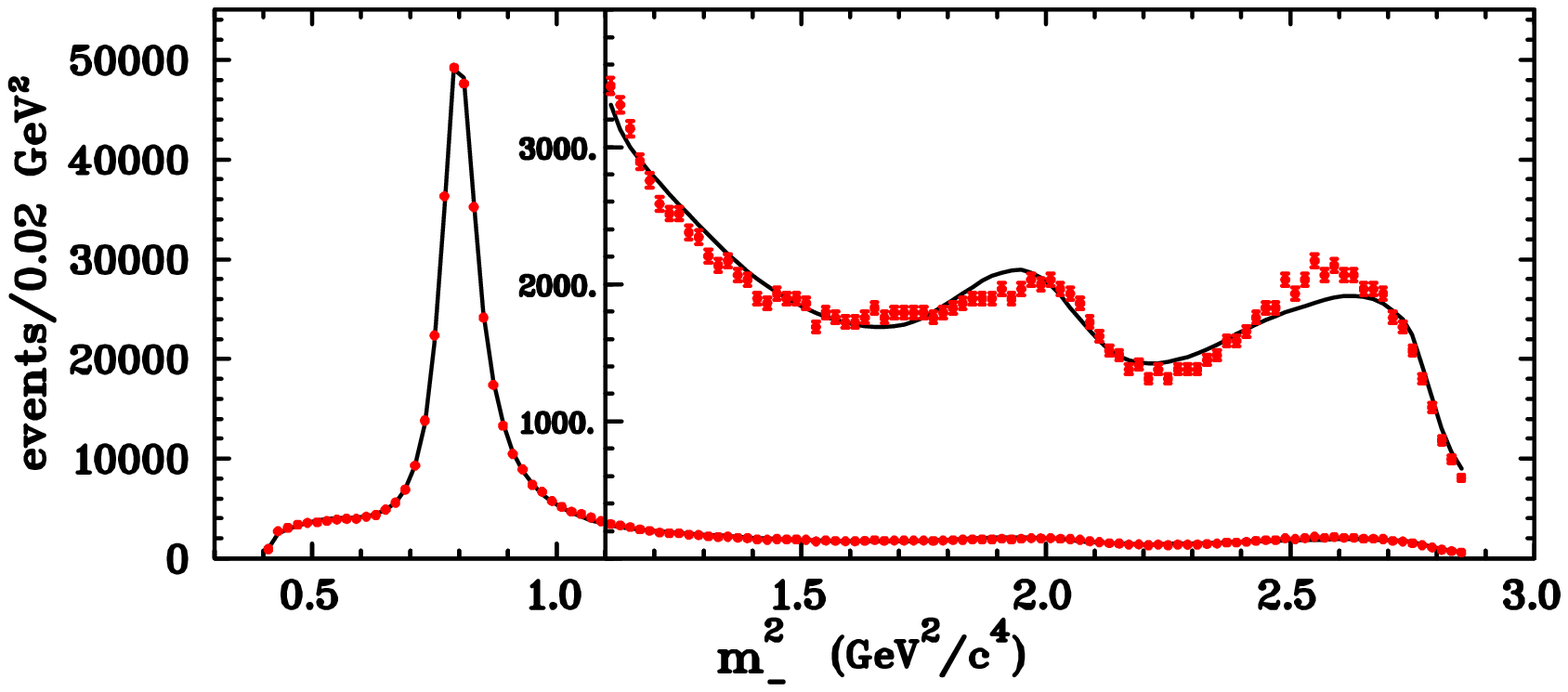}
\includegraphics[scale=0.35]{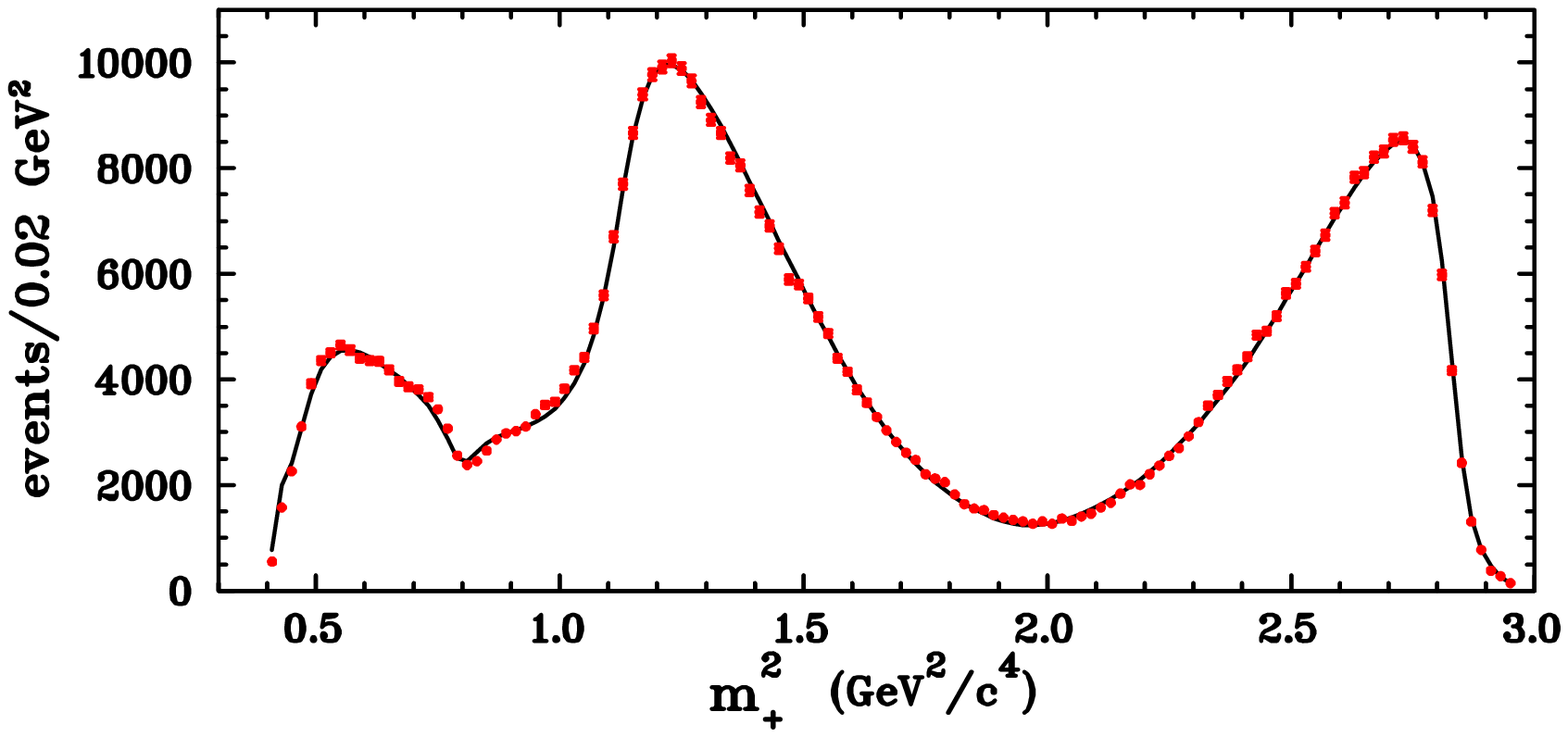}
\includegraphics[scale=0.36]{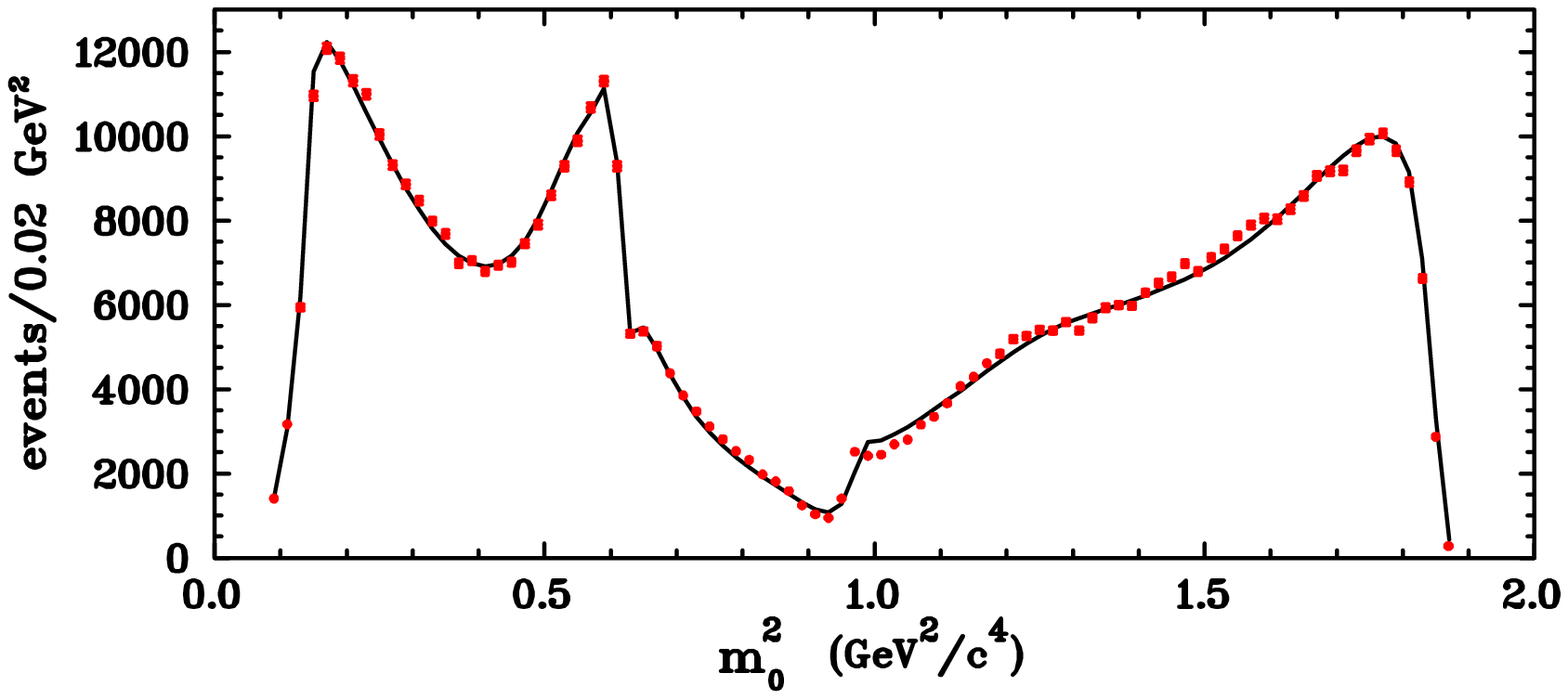}
\caption{Result of the present fit compared to  the Dalitz-plot projection of the Belle data~\cite{ZhangPRL99}}
\label{Belle}       % Give a unique label
\end{center}
\end{figure}

 The present model has 28 free parameters, mostly unknown transition form factors and  a minimization procedure is used to reproduce the $K^0_S \pi^-, K^0_S \pi^+$ and $\pi^+ \pi^-$ squared effective mass $m_\pm^2=s_\pm$ and $m_0^2=s_0$ projections of the experimental Dalitz plot analyses~\cite{SanchezPRL105,ZhangPRL99}.
Preliminary results are shown in Fig.~\ref{Belle} for  the Belle Collaboration analysis~\cite{ZhangPRL99}.
Similar results are obtained with the BABAR analysis~\cite{SanchezPRL105}.

\paragraph{Conclusion}
This theoretically  constrained analysis might be useful to improve the determination of the  $D^0$-$\bar D^0$ mixing parameters and of the CKM angle $\gamma$.

%Text with citations \cite{Ref1} and \cite{Ref2}.
%\subsection{Subsection title}
%\label{sec:2}

%Don't forget to give each section and subsection a unique label (see Sect.~\ref{sec:1}).

%\paragraph{Paragraph headings} Use paragraph headings as needed.
%\begin{equation}
%\end{equation}

%\begin{figure*}
%%  \includegraphics[width=0.75\textwidth]{fb20_logo.eps}
%\end{figure*}
%\begin{table}
%\begin{tabular}{lll}
%\hline\noalign{\smallskip}
%\end{tabular}
%\end{table}

%\begin{acknowledgements}
%\end{acknowledgements}



\end{document}